\documentclass[english,aps,pre,reprint,showpacs,superscriptaddress,floatfix]{revtex4-1}
\usepackage[T1]{fontenc}
\usepackage[latin9]{inputenc}
\usepackage{color}
\usepackage{babel}
\usepackage{amsmath}
\usepackage{amssymb}
\usepackage{graphicx}
\usepackage[unicode=true,pdfusetitle,
 bookmarks=true,bookmarksnumbered=false,bookmarksopen=false,
 breaklinks=true,pdfborder={0 0 0},backref=false,colorlinks=true]
 {hyperref}
\usepackage{breakurl}

\makeatletter

\providecommand{\tabularnewline}{\\}

\@ifundefined{textcolor}{}
{%
 \definecolor{BLACK}{gray}{0}
 \definecolor{WHITE}{gray}{1}
 \definecolor{RED}{rgb}{1,0,0}
 \definecolor{GREEN}{rgb}{0,1,0}
 \definecolor{BLUE}{rgb}{0,0,1}
 \definecolor{CYAN}{cmyk}{1,0,0,0}
 \definecolor{MAGENTA}{cmyk}{0,1,0,0}
 \definecolor{YELLOW}{cmyk}{0,0,1,0}
 }

\makeatother

\begin{document}

\title{Reconstruction of network structures from repeating spike patterns
in simulated bursting dynamics}

\author{Hao Song}

\affiliation{Institute of Physics, Academia Sinica, Nankang, Taipei, Taiwan 115,
R.O.C.}

\author{Chun-Chung Chen}

\email{cjj@phys.sinica.edu.tw}

\affiliation{Institute of Physics, Academia Sinica, Nankang, Taipei, Taiwan 115,
R.O.C.}

\author{Jyh-Jang Sun}

\affiliation{Neuro-Electronics Research Flanders, Kapeldreef 75, 3001 Leuven,
Belgium}

\author{Pik-Yin Lai}

\affiliation{Institute of Physics, Academia Sinica, Nankang, Taipei, Taiwan 115,
R.O.C.}

\affiliation{Department of Physics and Center for Complex Systems, National Central
University, Chungli, Taiwan 320, R.O.C.}

\author{C. K. Chan}

\affiliation{Institute of Physics, Academia Sinica, Nankang, Taipei, Taiwan 115,
R.O.C.}

\affiliation{Department of Physics and Center for Complex Systems, National Central
University, Chungli, Taiwan 320, R.O.C.}

\date{\today}
\begin{abstract}
Repeating patterns of spike sequences from a neuronal network have
been proposed to be useful in the reconstruction of the network topology.
Reverberations in a physiologically realistic model with various physical
connection topologies (from random to scale-free) have been simulated
to study the effectiveness of the pattern-matching method in the reconstruction
of network topology from network dynamics. Simulation results show
that functional networks reconstructed from repeating spike patterns
can be quite different from the original physical networks; even global
properties, such as the degree distribution, cannot always be recovered.
However, the pattern-matching method can be effective in identifying
hubs in the network. Since the form of reverberations are quite different
for networks with and without hubs, the form of reverberations together
with the reconstruction by repeating spike patterns might provide
a reliable method to detect hubs in neuronal cultures.
\end{abstract}


\maketitle

\section{Introduction}

 One of the most fundamental problems in the understanding of neural
networks is to relate the observed dynamics of a network to its connection
structure \cite{feldt2011dissecting}. Since networks made of similar
elements and interactions, such as our brains, can perform seemingly
very different tasks, it is believed that the functions of the network
are mainly governed by its connection structure not its constituents.
In general, the behavior of a network is governed by the dynamics
of its individual elements, their interactions and their connection
topology \cite{albert2002statistical}. It is straight forward to
compute the dynamics of the network when all these three factors are
known. However, the reverse problem of reconstructing the structure
of the network from its dynamics is highly non-trivial \cite{ching2013extracting}.
It is possible that different physical connection structures might
give rise to similar observed dynamics. Also, the dynamics of a network
can be history dependent (memory) without any changes in physical
connections.

 However, since the physical connection of a neural network is usually
not available and the network dynamics is the only information one
can obtain from a neural network, many studies have been devoted to
the studies of network reconstruction from the observed dynamics;
such as cross correlation \cite{pires2014modeling}, Granger causality
\cite{granger1969investigating}, transfer entropy \cite{schreiber2000measuring}
etc. Reconstructing the underlying network connection solely from
the measurement of the time-series signal of the elements is in general
a difficult task, albeit it can be achieved recently for undirected
networks of uniform coupling strengths \cite{ching2013extracting}
or for directed networks with loop-free structure \cite{pires2014modeling}.
The problem becomes even more challenging for the case of neuronal
networks due to the directed synaptic connections and their dynamically
dependent plasticity. One intuitively simple attempt for neuronal
network reconstruction is based on repeating spatial temporal firing
patterns \cite{sun2010selforganization} of the network. In the concept
of cell assembly \cite{hebb1949theorganization}, proposed by Hebb
as a possible mechanism for the realization of higher brain functions
from interconnected neurons, these repeating spike patterns are believed
to be related to the network connection structure of the cell assembly.
One of the earliest attempts for the search of cell assembly \cite{abeles1988detecting}
was to use multi-site recordings of neuronal activities to examine
the spatial--temporal firing patterns in cultures and \emph{in vivo}
experiments. In these studies, the repeating spike patterns are considered
fundamental because they represent a dynamical characteristic of the
network which should be intimately related to its connection structure.
Although the relation between these repeating spike patterns and the
network structure is still far from clear, one can still ``reconstruct''
network structures from these pattern, if the notion of ``fire together,
wire together'' is used. With this heuristic rule, repeating spike
patterns obtained from neuronal cultures have been used to reconstruct
network structures \cite{sun2010selforganization}. 
\begin{figure*}
\includegraphics[width=0.24\textwidth]{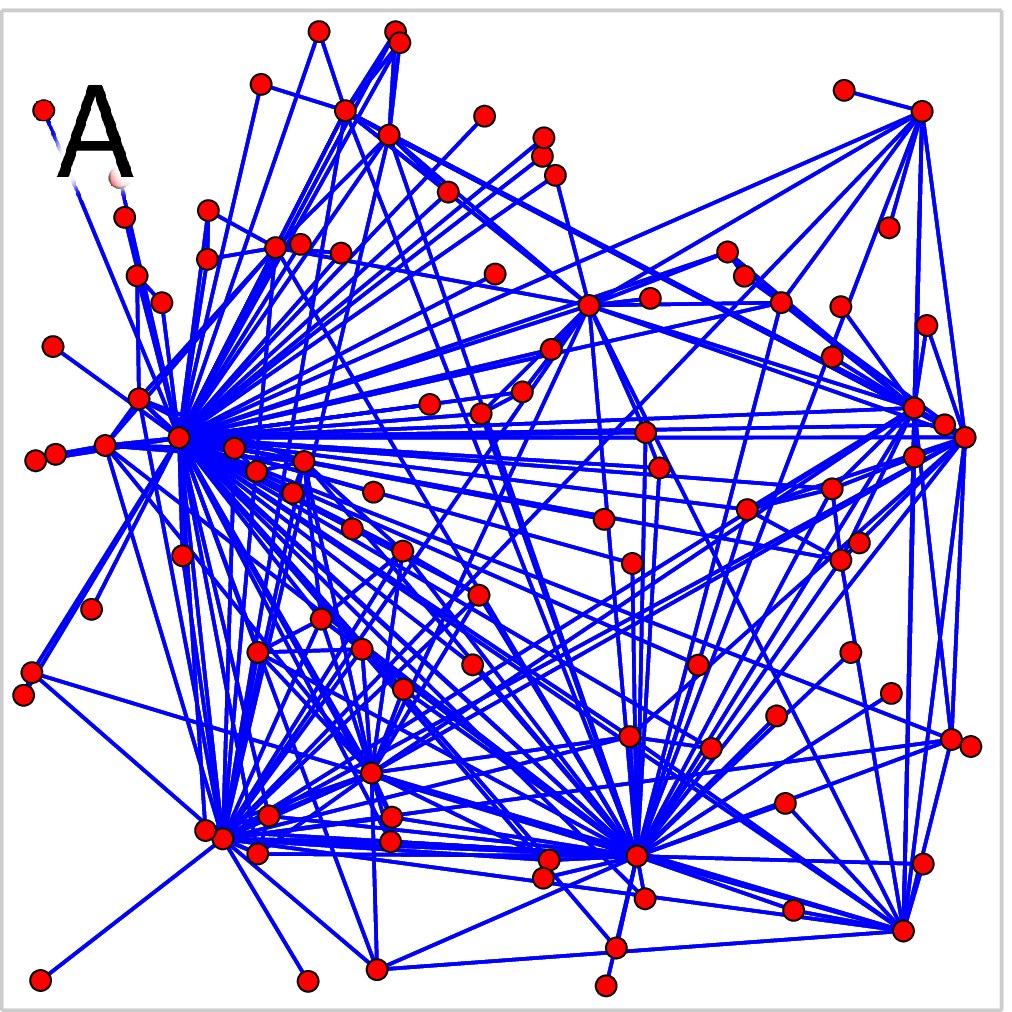}\includegraphics[width=0.24\textwidth]{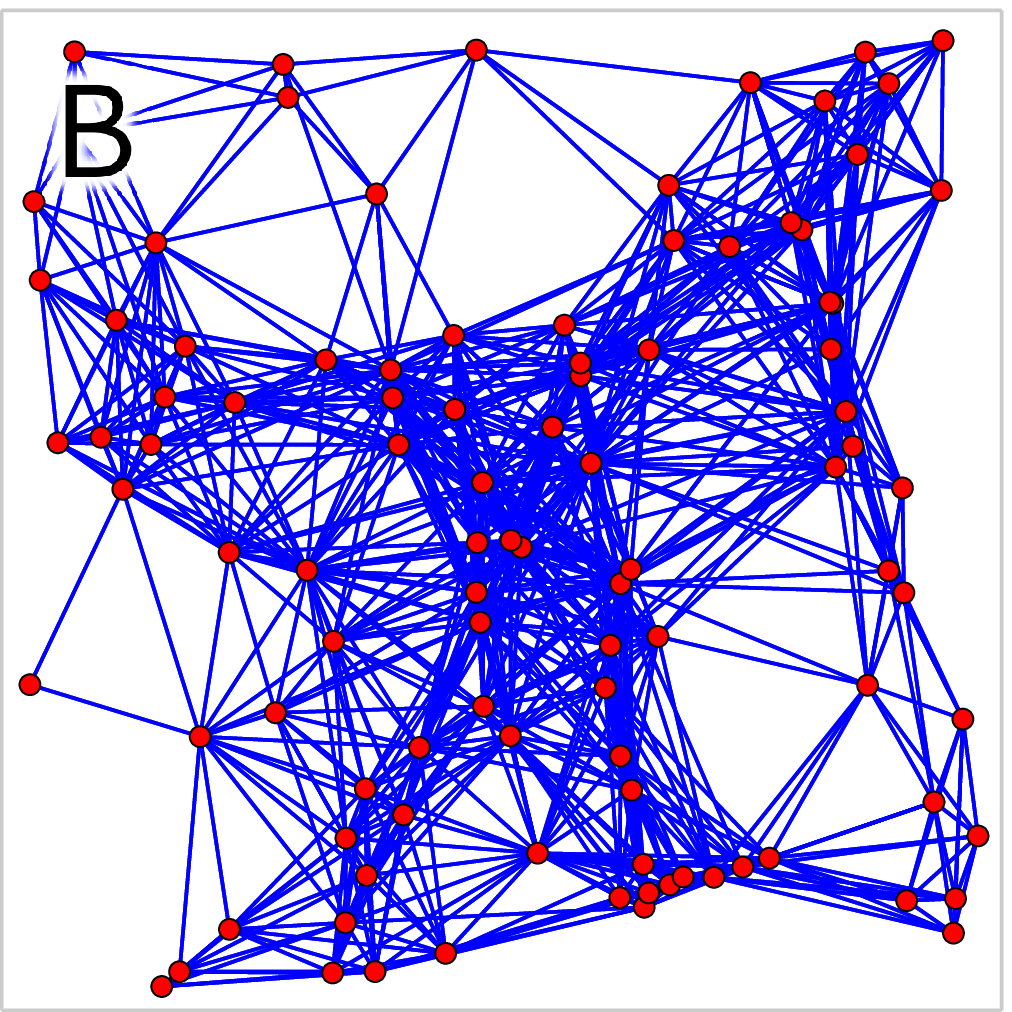}\includegraphics[width=0.24\textwidth]{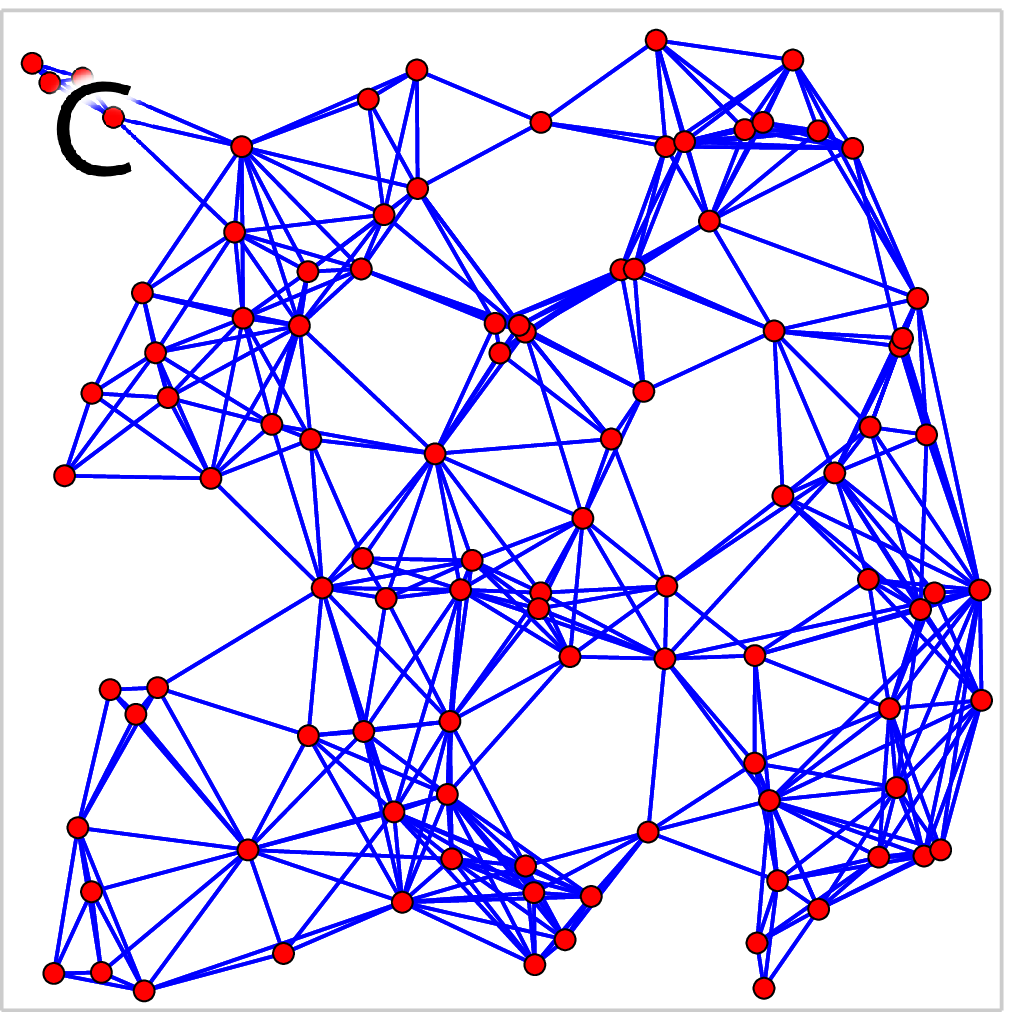}\includegraphics[width=0.24\textwidth]{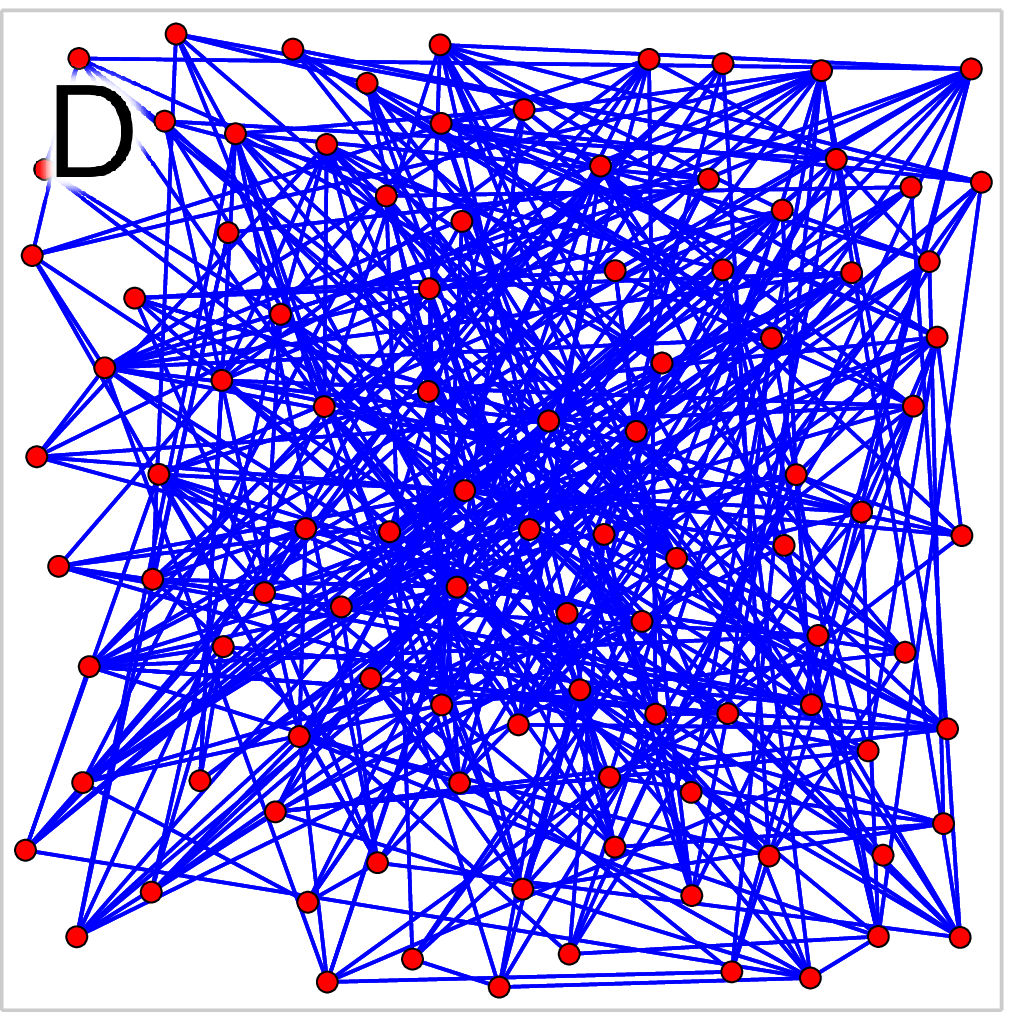}\\
\includegraphics[width=0.24\textwidth]{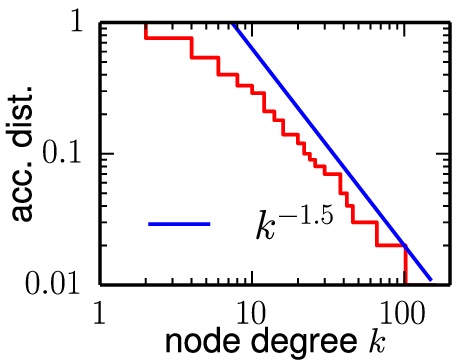}\includegraphics[width=0.24\textwidth]{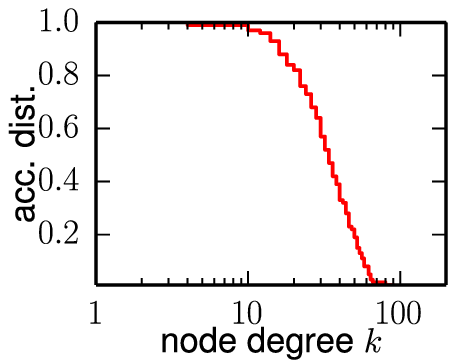}\includegraphics[width=0.24\textwidth]{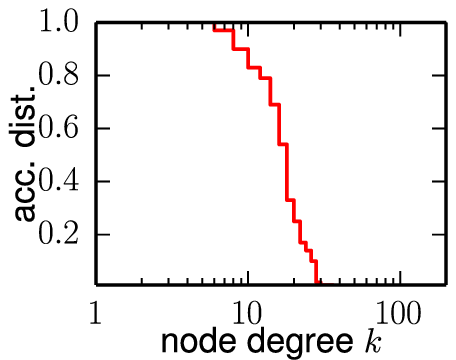}\includegraphics[width=0.24\textwidth]{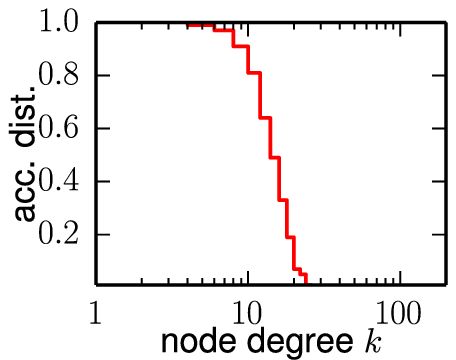}

\protect\caption{(Color online) Different network structures generated with Morita's
method \cite{morita2006crossovers} with A: $\gamma=1.5$, $\delta=2.7$;
B: $\gamma=3$, $\delta=4.7$; C: $\gamma=5$, $\delta=2$; and a
random network D with similar degree distribution to C. \label{fig:Different-network-structures}}
\end{figure*}

 One important characteristics of network reconstructed from dynamics
is that the reconstructed network structures are only ``functional''.
Since it is known that even a simple network of cultured neurons can
display different kind of dynamics within a short time when drugs
\cite{maeda1995themechanisms} are added to the perfusion. The functional
network deduced from dynamics might not reflect the physical connections
in the system. It is not uncommon that different studies of functional
networks in the brain can find both scale-free \cite{eguiluz2005scalefree}
and small-world \cite{achard2006aresilient} topologies. Presumably,
different kinds of functional topologies could be created dynamically
from the same physical network. Very little is known about the relation
between the functional network and its physical counter part. It would
be desirable if there were control experiments in which both the physical
connection and resultant dynamics are known so that we can study the
relation between the the functional network and the physical network.
Unfortunately, these ideal experiments do not exist yet. As a remedy
for such a situation, computer simulations might be useful. In this
article, we describe our studies on the relation between the physical
network and its reconstructed functional network from repeating spike
patterns in the simulation of a physiologically realistic model \cite{volman2007calcium}
which is known to reproduce reverberations found in experiments \cite{lau2005synaptic}.
Our results suggest that the method of repeating spike patterns might
be useful for the local reconstruction of hubs during reverberation
if hubs exist. However, as global structure is concerned, such as
the degree distribution, the repeating-spike-pattern method might
not be applicable.

\section{Methods}

Our simulation studies of relation between physical networks and function
networks are consisted of three main steps; namely a) network generation,
b) simulation of network dynamics and c) functional-network reconstruction.
We use a network generation method in which the topologies of the
network can be changed continuously from random to scale free by tuning
some system parameters. This flexibility will allow us to use the
same structure generation algorithm for various topologies and more
importantly to generate structures in between random and scale free.
For the network dynamic simulation, a physiologically realistic model
is chosen. The model is known to be able to generate reverberation
patterns similar to experimental observations in neuronal cultures
\cite{lau2005synaptic}. As for the functional network reconstruction,
we use the repeating-spike-pattern method which have been used to
identify network topologies in neuronal cultures. Details of these
three steps are given below.

\subsection{Network Generation\label{sub:Network-Generation}}

In order to generate networks of different structures, we used the
method described by Morita \cite{morita2006crossovers,gosak2010optimal}.
In this method, $N$ vertices of the network are distributed randomly
on a 2D unit square. Two vertices $i$ and $j$ are connected if their
locations satisfy the equation, 
\begin{equation}
\frac{2l_{i,j}^{2}}{a_{i}a_{j}}<\delta,\label{eq:link}
\end{equation}
where $l_{i,j}$ is the distance between the two points, $\delta$
is a threshold parameter, and $a_{i}=\left(i/N\right)^{1/\left(1-\gamma\right)}$
for a given parameter $\gamma>1$. A main advantage of Morita's method
is that different types of network topology can be generated by continuous
changes of the parameters $\gamma$ and $\delta$. Figure~\ref{fig:Different-network-structures}
shows three network structures generated with different parameters
in Morita's method along with a Erd\H{o}s--R\'{e}nyi random network \cite{erdhos1959onrandom}
as well as their corresponding accumulative degree ($k$) distributions.
From the left, the degree distributions of the networks in Fig.~\ref{fig:Different-network-structures}
range from scale free with a power of $-2.5$ to Gaussian-like similar
to the random network. Beside being tunable in their degree distributions,
the networks generated with Morita's method are also pertinent to
a 2D geometry, similar to a neural culture, which is apparent when
comparing the last (third from the left) network structure to the
random network with similar degree distribution on the right of Fig.~\ref{fig:Different-network-structures}:
The vertices that are closer to each other in the geometrical space
are more likely to be connected.

\begin{figure}
\begin{centering}
\includegraphics[width=0.5\textwidth]{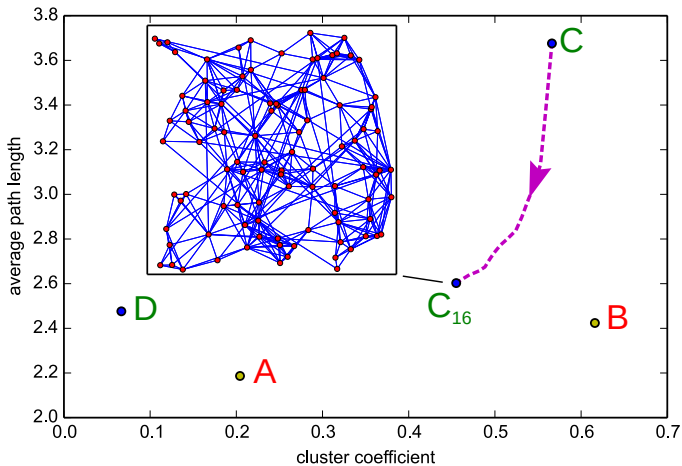}
\par\end{centering}

\protect\caption{(Color online) Average path lengths versus cluster coefficients for
the networks A to D as shown in Fig. \ref{fig:Different-network-structures}.
Network ${\rm C}_{16}$, as shown in inset, is network C with 16 pairs
of links rewired to minimize average path length while preserving
the degree structure for all the nodes. It represents a small-world
construction from the geometrically constrained network C.\label{fig:Average-path-lengths}}
\end{figure}
Following recent findings of small-world \cite{watts1998collective}
structures in cultured networks \cite{downes2012emergence,desantos-sierra2014emergence},
we calculate the cluster coefficients as well as average path length
between nodes for the networks in Fig. \ref{fig:Different-network-structures}
and plot them as labeled in Fig. \ref{fig:Average-path-lengths}.
Similar to the lattice based regular network in \cite{watts1998collective},
the network C has a large cluster coefficient and relatively long
average path length since it is similarly constrained to a geometric
space while connections are short-ranged. We perform pair-rewiring
steps similar to that described in \cite{maslov2002specificity} to
minimize the average path length of C while preserving its degree
distribution. The network ${\rm C}_{16}$ as depicted in the inset
of Fig. \ref{fig:Average-path-lengths}, arrived after 16 pair-rewiring
steps, has an average path length on par with random network D while
maintaining about 80\% of its cluster coefficient. We consider network
${\rm C}_{16}$ as a small-world intermediate between networks C and
D.

In the following analysis, we will limit our consideration to the
five different networks, each of 100 neurons, as shown in Fig.~\ref{fig:Different-network-structures}
and Fig. \ref{fig:Average-path-lengths} with bidirectional connectivity.

\subsection{Model of network dynamics\label{sub:Neuron-and-Synaptic}}

\noindent The notion of ``fire together wire together'' has been
used in many studies to identify functional connection in cultures.
Usually, ``fire together'' can mean strong correlation or even synchronization.
It is well known that neuronal cultures develop synchronized bursting
during development although its origin is still not clear. In a study
of using core patterns to identify connection during development,
Sun \emph{et al} \cite{sun2010selforganization} have also used the
repeated spike patterns in synchronized bursts to identify connections.
Here, our simulation is based on a recent model by Volman \emph{et
al} \cite{volman2007calcium} describing the dynamics of reverberation
activity in cultured neural networks in which synchronized bursts
have been observed. In this model, the dynamics of neurons follows
the Morris--Lecar model \cite{morris1981voltage,rinzel1989analysis},\begin{subequations}\label{sub:Morris-Lecar}
\begin{eqnarray}
C\frac{dV}{dt} & = & -I_{{\rm ion}}+G\left(V_{r}-V\right)+I_{{\rm bg}},\\
\frac{dW}{dt} & = & \theta\frac{W_{\infty}-W}{\tau_{W}}
\end{eqnarray}
\end{subequations}where 
\begin{equation}
I_{{\rm ion}}=g_{{\rm Ca}}m_{\infty}\left(V-V_{{\rm Ca}}\right)+g_{{\rm K}}W\left(V-V_{{\rm K}}\right)+g_{{\rm L}}\left(V-V_{{\rm L}}\right)
\end{equation}
is the current through the membrane ion channels,\begin{subequations}
\begin{eqnarray}
\tau_{W} & = & \left(\cosh\frac{V-V_{3}}{2V_{4}}\right)^{-1},\\
W_{\infty} & = & \frac{1}{2}\left(1+\tanh\frac{V-V_{3}}{V_{4}}\right),\\
m_{\infty} & = & \frac{1}{2}\left(1+\tanh\frac{V-V_{1}}{V_{2}}\right)
\end{eqnarray}
\end{subequations}are the voltage dependent dynamic parameters, and
the threshold $V_{{\rm th}}$ of membrane potential defines the spiking
events which result in synchronous releases of neural transmitters
at the efferent synapses. Additionally, a residual calcium variable
$R_{{\rm Ca}}$ driven by the spiking events,
\begin{equation}
\frac{d}{dt}R_{{\rm Ca}}=\frac{-\beta R_{{\rm Ca}}^{n}}{k_{R}^{n}+R_{{\rm Ca}}^{n}}+I_{p}+S\gamma\log\frac{R_{{\rm Ca}}^{0}}{R_{{\rm Ca}}},
\end{equation}
where the spike train $S=\sum_{\sigma}\delta\left(t-t_{\sigma}\right)$
with $t_{\sigma}$ being the time of the spike event $\sigma$, is
used to determine the rate,
\begin{equation}
\eta=\eta_{{\rm max}}\frac{R_{{\rm Ca}}^{m}}{k_{a}^{m}+R_{{\rm Ca}}^{m}},\label{eq:asynchro-rate}
\end{equation}
of synapse-dependent asynchronous releases following an independent
Poisson process at each efferent synapse. The neural transmitters
released by the spike-driven synchronous and calcium-dependent asynchronous
events follow a four-state decaying dynamics based on a modification
of the Tsodyks--Uziel--Markram (TUM) model \cite{tsodyks2000synchrony},\begin{subequations}\label{sub:TUMs-dynamics}
\begin{eqnarray}
\frac{dX}{dt} & = & \frac{Q}{\tau_{s}}+\frac{Z}{\tau_{r}}-uXS-X\xi\\
\frac{dY}{dt} & = & -\frac{Y}{\tau_{d}}+uXS+X\xi\label{eq:active-state}\\
\frac{dZ}{dt} & = & \frac{Y}{\tau_{d}}-\frac{Z}{\tau_{r}}-\frac{Z}{\tau_{l}}\\
\frac{dQ}{dt} & = & \frac{Z}{\tau_{l}}-\frac{Q}{\tau_{s}}.\label{eq:super-inactive-state}
\end{eqnarray}
\end{subequations}where $\xi=\bar{\xi}\sum_{a}\delta\left(t-t_{a}\right)$
summing over asynchronous release event $a$ with Poisson rate given
by (\ref{eq:asynchro-rate}), to include a super-inactive state $Q$
\cite{volman2007calcium}. Multiplying by the synaptic weights, the
fractions of neural transmitters in the active state $Y$ (\ref{eq:active-state})
determine the contribution of the afferent synapses to the membrane
conductance $G$ of a post-synaptic neuron through a linear sum 
\begin{equation}
G_{i}=\sum_{j}w_{ji}Y_{ji}
\end{equation}
over all pre-synaptic neurons $j$ of a given post-synaptic neuron
$i$. Following \cite{volman2007calcium}, the synaptic weights $w$
are randomly drawn from a truncated Gaussian distribution with a width
that is $\pm20\%$ of its mean $\bar{w}$. The super-inactive state
$Q$ (\ref{eq:super-inactive-state}) of neural transmitters plays
a part in taking up neural transmitters during a burst of reverberations
and eventually terminates the reverberatory burst. 
\begin{figure}
\begin{centering}
\includegraphics[width=0.5\textwidth]{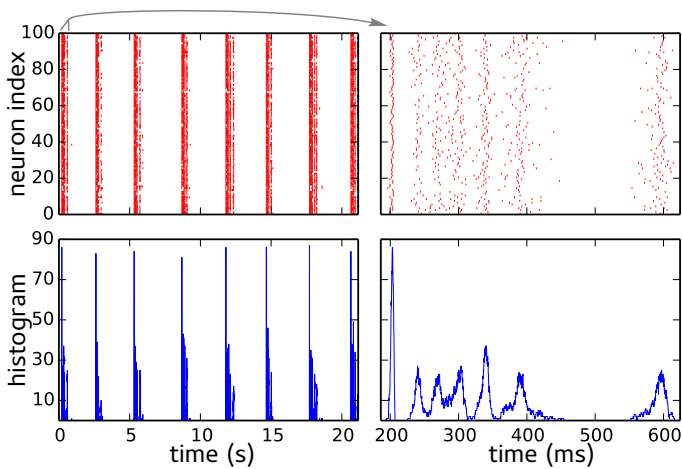}
\par\end{centering}

\protect\caption{(Color online) Typical bursting behavior of a random network (D in
Fig.~\ref{fig:Different-network-structures}) with neuron and synaptic
transmission models described in \cite{volman2007calcium}. The left
panels show eight bursts for a duration of about 20 seconds while
the right panels are the zoom-in on the first burst. The upper panels
are the raster plots marking the spiking events for the neurons indexed
on the vertical axes. The bottom panels are histograms with the vertical
axis showing the number of spikes within a sliding bin of 4 ms. \label{fig:Bursting-behavior-of}}
\end{figure}
 In this model, there are two types of noises. The first one is the
background current $I_{{\rm bg}}$ which is uniformly added to every
neuron to mimic a noisy environment. The other is the asynchronous
release due to the residual calcium in the neuron after firing. The
latter types of noise can be considered a kind of short term memory
because it is related to the firing history of the neuron. The Poisson
processes modeling these asynchronous events are the only sources
of stochasticity in the model. A typical bursting state of the system
is shown in Fig.~\ref{fig:Bursting-behavior-of} with sub-burst reverberations.
Notice that in \cite{volman2007calcium}, the reverberatory burst
was triggered by an external stimulus while spontaneous bursting is
possible but rare. This corresponds to the regime of high to infinite
restitution ratio in our cases where the networks will be silent by
themselves.

Note that the chosen residual calcium dynamics and the synaptic mechanisms
are responsible for the reverberation (bursts) shown in Fig.~\ref{fig:Bursting-behavior-of}.
The dynamics of the network depends on the parameters of the model
as well as the network topology. As there are a total of 30 parameters
used from (\ref{sub:Morris-Lecar}) to (\ref{sub:TUMs-dynamics})
to define the model in \cite{volman2007calcium}, we do not intend
to be comprehensive in exploring the entire phase space. 
\begin{table}
\protect\caption{Values of fixed model parameters\label{tab:volman-model-parameters}}

\begin{centering}
\begin{tabular*}{0.8\columnwidth}{@{\extracolsep{\fill}}rlrlrl}
\hline 
\multicolumn{6}{c}{Morris--Lecar model}\tabularnewline
\hline 
$V_{{\rm Ca}}$ & $100$ ${\rm mV}$ & $V_{2}$ & $15$ ${\rm mV}$ & $g_{{\rm L}}$ & $0.5$ ${\rm mS}$\tabularnewline
$V_{{\rm K}}$ & $-70$ ${\rm mV}$ & $V_{3}$ & $0$ ${\rm mV}$ & $C$ & $1$ ${\rm \mu F}$\tabularnewline
$V_{{\rm L}}$ & $-65$ ${\rm mV}$ & $V_{4}$ & $30$ ${\rm mV}$ & $\text{\ensuremath{\theta}}$ & $0.2$ ${\rm ms}^{-1}$\tabularnewline
$V_{r}$ & $0$ ${\rm mV}$ & $g_{{\rm Ca}}$ & $1.1$ ${\rm mS}$ & $V_{{\rm th}}$ & $10$ ${\rm mV}$\tabularnewline
$V_{1}$ & $-1$ ${\rm mV}$ & $g_{{\rm K}}$ & $2$ ${\rm mS}$ &  & \tabularnewline
\hline 
\end{tabular*}
\par\end{centering}

\begin{centering}
\begin{tabular*}{0.8\columnwidth}{@{\extracolsep{\fill}}rlrlrl}
\hline 
\multicolumn{6}{c}{TUM synaptic transmission}\tabularnewline
\hline 
$\tau_{d}$ & $10$ ${\rm ms}$ & $\tau_{l}$ & $600$ ${\rm ms}$ & $u$ & $0.2$\tabularnewline
$\tau_{r}$ & $300$ ${\rm ms}$ & $\tau_{s}$ & $5000$ ${\rm ms}$ & $\bar{\xi}$ & $0.02$\tabularnewline
\hline 
\end{tabular*}
\par\end{centering}

\centering{}%
\begin{tabular*}{0.8\columnwidth}{@{\extracolsep{\fill}}rlrlrl}
\hline 
\multicolumn{6}{c}{Residual calcium dynamics}\tabularnewline
\hline 
$\beta$ & $0.005$ $\frac{{\rm \mu M}}{{\rm ms}}$ & $\gamma$ & $0.033$ & $k_{a}$ & $0.1$ ${\rm \mu M}$\tabularnewline
$k_{R}$ & $0.4$ ${\rm \mu M}$ & $R_{{\rm Ca}}^{0}$ & $2000$ ${\rm \mu M}$ & $m$ & $4$\tabularnewline
$I_{p}$ & $1.1\times10^{-4}$ $\frac{{\rm \mu M}}{{\rm ms}}$ & $\eta_{{\rm max}}$ & $0.32$ ${\rm ms}^{-1}$ & $n$ & $2$\tabularnewline
\hline 
\end{tabular*}
\end{table}
 Instead, we fix all parameters with values shown in Table~\ref{tab:volman-model-parameters},
similar to what were documented in \cite{volman2007calcium}, except
for the background current $I_{{\rm bg}}$ and mean synaptic weight
$\bar{w}$, which are varied to obtain a network with spontaneous
bursting behavior. Since the reverberations reproduced in this model
with a random network are consistent with those found in experiments
\cite{volman2007calcium}, presumably the simulations we carry out
here with different topologies are physiologically realistic.

\subsection{Repeating-spike-pattern detection}

We follow the method in \cite{sun2010selforganization} to use repeating
spike patterns for the reconstruction of the network in our simulation.
Briefly, repeating spike patterns are defined as repetitive patterns
of a sequence of firings from different neurons and a link is assigned
between two neurons (wire together) if the spikes of these two neurons
are linked in time (fire together). To detect repeating spike patterns,
a template-matching algorithm following \cite{rolston2007precisely}
were implemented to locate the repetitive firing patterns among these
spiking data to form the repeating patterns.

Since the repeating spike patterns are assumed to have linear connections
with only links between adjacent neurons in the spike-time sequence\cite{sun2010selforganization},
this reconstruction method ignores the possibility of ``branching''
in the propagation of spiking activity and could likely predict some
connections that were not present in the actual network structure.
Note that in \cite{sun2010selforganization}, spike sorting was needed
to produce a vector of spike times for each identified neuron from
the time series recorded by a multi-electrode array. In our case,
since all the neurons are known, no spike sorting is needed and only
spike detection is performed to produce the spike-time vector. In
the current study, a similar MATLAB code as that was used in \cite{sun2010selforganization}
is applied to the spiking data generated by the dynamics in \ref{sub:Neuron-and-Synaptic}
on the networks described in \ref{sub:Network-Generation} to find
repeating spike patterns for the systems.

\section{Results}

With the methods described above, it is obvious that for a network
with fixed topology, the dynamics of the network can be strongly dependent
on the choice of synaptic strength and the background current. If
we want to reconstruct the topology of the network from the dynamics,
we would like to create a situation in which the simulation results
will be similar to those in experiments. It is known that synchronized
bursting activities emerge as cortical neuronal cultures develop and
the network connectivity increases \cite{jia2004connectivities,lai2006growthof}.
As mentioned earlier, for a developing neural cultures, repeating
spike patterns will appear during spontaneous bursting. Thus, our
goal is to reproduce spontaneous bursting as in \cite{sun2010selforganization}.
In the original model, the background current is chosen in such a
way that the network is in a quiescent steady state in the absence
of input but would produce a single, reverberatory burst after an
excitatory current is presented to one of the neuron in the network
\cite{volman2007calcium}. Since we would like to compare with \cite{sun2010selforganization},
we need to adjust the background current to give spontaneous, synchronized
bursting. Such spontaneous burst will be followed by a quiescent state.
In experiments, the durations of the quiescent state decrease as the
culture mature while the durations of the bursting also decrease but
less significantly. In the simulation, we can use the ratio between
these two states as an indication of the age of the culture. For a
fixed background current, this ratio can be tuned by the synaptic
strength.

To define the bursting state of the network in our simulation, we
use a hysteretic criteria:\label{hysteretic criteria} the system
is considered to enter bursting/resting state when there have been
more/less than nine/three distinct neurons firing spikes within last
200 ms time window. These numbers are chosen to be similar to those
in real experiments. With the defined states, we can measure the durations
of bursting and resting episodes. In simulations, for a given background
current $I_{{\rm bg}}$ ranging from $27.0$ to $29.5$ $\mu$A, we
adjust the the mean synaptic weight $\bar{w}$ of the network to obtain
bursting behavior similar to what is shown Fig.~\ref{fig:Bursting-behavior-of}.
\begin{figure}
\begin{centering}
\includegraphics[width=0.5\textwidth]{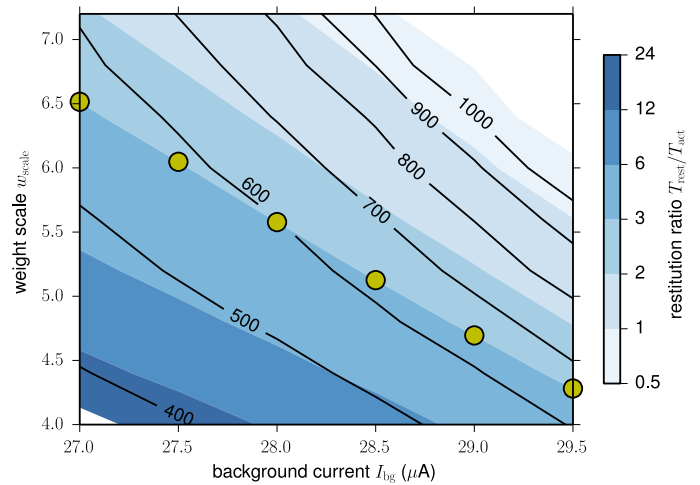}
\par\end{centering}

\protect\caption{(Color online) Phase diagram for bursting dynamics on a random network.
Colors (shades of gray) indicate the restitution ratio of bursting
while labeled lines show the average bursting period in milliseconds.\label{fig:Phase-diagram-for}}
\end{figure}
 For further pattern matching, we choose the mean synaptic weight
so that the ratio between average durations of resting to bursting
states (restitution ratio) is 3 to 1 \cite{ratio} as indicated by
circles in Fig.~\ref{fig:Phase-diagram-for}. Since $I_{{\rm bg}}$
determines how easily a neuron is ready to fire while $w_{{\rm scale}}$
determines how strongly a neuron's firing can influence others, one
would expect an increase in $I_{{\rm bg}}$ would be compensated by
a decrease of $w_{{\rm scale}}$. From our simulations, this trend
can be more or less observed for networks with more uniform, Gaussian
degree distribution (network C and D in Fig.~\ref{fig:Different-network-structures}
as well as ${\rm C_{16}}$ in Fig.~\ref{fig:Average-path-lengths}).
In the following, we report the results of our simulation with the
parameters given in Table~\ref{tab:volman-model-parameters} while
the background current $I_{{\rm bg}}$ and mean synaptic weight scale
$w_{{\rm scale}}$ being adjusted to give the restitution ratio of
3 to 1.

\subsection{Bursting behavior}

For each of the networks given in Fig.~\ref{fig:Different-network-structures},
we perform simulations to obtain spontaneous bursting by varying the
background current $I_{{\rm bg}}$ and mean synaptic weight scale
$w_{{\rm scale}}$. 
\begin{figure}
\begin{centering}
\includegraphics[width=0.5\textwidth]{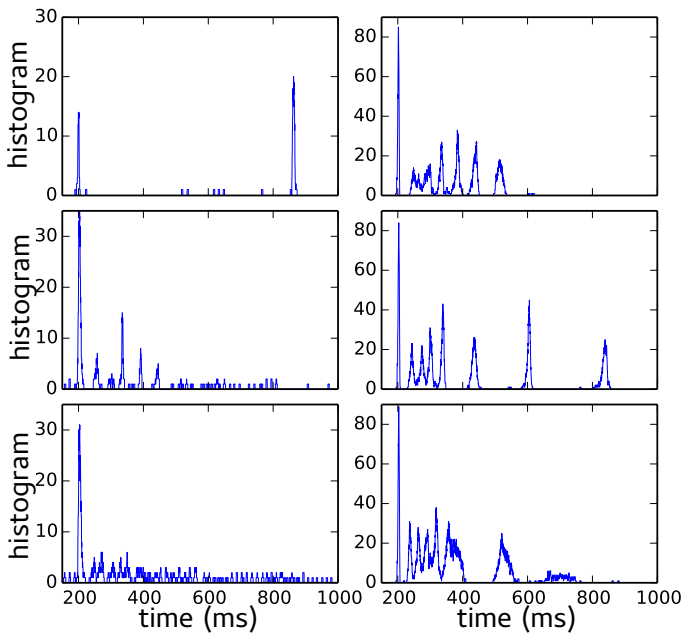}
\par\end{centering}

\protect\caption{(Color online) Comparison of bursting behavior for scale-free network
A (left) with random network D (right) in Figure \ref{fig:Different-network-structures}
at three different background current: from top down $I_{{\rm bg}}$
= $27.5$, $28.5$, $29.5$ $\mu A$. The horizontal axes measure
time in milliseconds while the vertical axes show the number of spikes
within a 4 ms sliding window.\label{fig:Comparison-of-bursting}}
\end{figure}
 Figure \ref{fig:Comparison-of-bursting} are the typical results
for spontaneous bursting for a scale-free and a random network. It
can be seen from Fig.~\ref{fig:Comparison-of-bursting} that the
main difference in the firing patterns between a random network and
a scale-free network is the reverberations within the synchronized
bursts of random network and their absence in the scale-free network.
This difference in bursting behavior is probably due to the existence
of hub neurons in the scale-free network which are found to stay active
the entire time for any reasonable choices of $I_{{\rm bg}}$ and
$w_{{\rm scale}}$ that allow for other neurons in the network to
be activated. This constant activation disrupts the quiescence between
the reverberations and breaks the coordinated firing of neurons required
by the reverberation. The situation is especially true for cases of
lower background current where stronger synaptic weights are required.
This is one of the reasons that we adopted the hysteretic criteria
described in \ref{hysteretic criteria} to define the bursting/resting
states of the systems.

\begin{figure}
\begin{centering}
\includegraphics[width=0.4\textwidth]{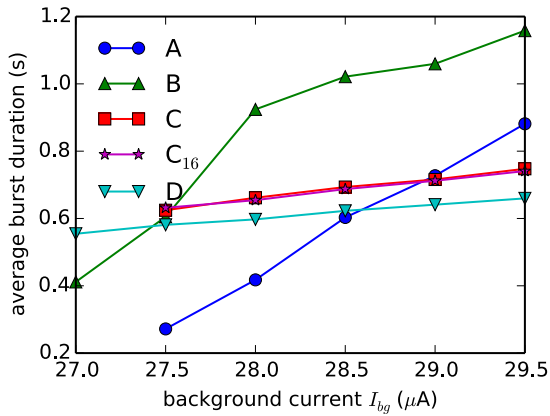}
\par\end{centering}

\protect\caption{(Color online) Average burst durations when mean synaptic weights
of a network is adjusted to have 3-to-1 rest-to-burst duration ratio
for the networks given in Fig.~\ref{fig:Different-network-structures}
and Fig.~\ref{fig:Average-path-lengths} as functions of the background
current given in the model. \label{fig:Average-burst-durations}}
\end{figure}
 The qualitative change of bursting behavior with different network
topologies can also be seen in the dependence of the mean burst duration
on $I_{{\rm bg}}$ as shown from Fig.~\ref{fig:Average-burst-durations}.
It can be seen in Fig.~\ref{fig:Average-burst-durations} that the
burst duration of networks with narrow degree distribution, such as
C, ${\rm C}_{16}$, and D, are insensitive to $I_{{\rm bg}}$ when
compared to networks with boarder degree distribution, such as A and
B. Beside the dependence of burst duration on $I_{{\rm bg}}$, we
note there are significantly less reverberations (repeated rises of
system activity level, as can be seen in left panel of Fig.~\ref{fig:Bursting-behavior-of})
within a burst for scale-free network A. It is less so for network
B where the reverberations are quite evident in the mid-range of background
current \cite{SuppMatt}. The insensitivity of burst duration on $I_{{\rm bg}}$
for networks C, ${\rm C}_{16}$, and D in Fig.~\ref{fig:Average-burst-durations}
shows the equivalence of excitability of neurons and efficacy of synapses
in networks of narrow degree distributions. Both of the networks also
retain clear reverberatory behavior for the entire range of background
current we considered \cite{SuppMatt}.

\subsection{Properties of repeating spike patterns}

With time series similar to those from Figure \ref{fig:Comparison-of-bursting}
for all the neurons in the networks, the template-matching method
in \cite{sun2010selforganization} is used to identify repeating spike
patterns. Data sets corresponding 8 bursting events of the networks
are used for repeating-spike-pattern identifications. This amounts
to about 3000 spikes per data set, similar to what was considered
in \cite{sun2010selforganization} limited by the computer capability
for the algorithm used. While more patterns will generally result
from a larger data set, we do not expect a qualitative change to our
conclusions presented below \cite{SuppMatt}. Each identified repeating
spike pattern is cross-checked with the original network used in the
simulation to see if: a) the nodes in the pattern form a connected
subnetwork with the original links and b) the nodes in the identified
repeating spike pattern are actually connected sequentially in the
nominal network as suggested in \cite{sun2010selforganization}. 
\begin{figure}
\begin{centering}
\includegraphics[width=0.5\textwidth]{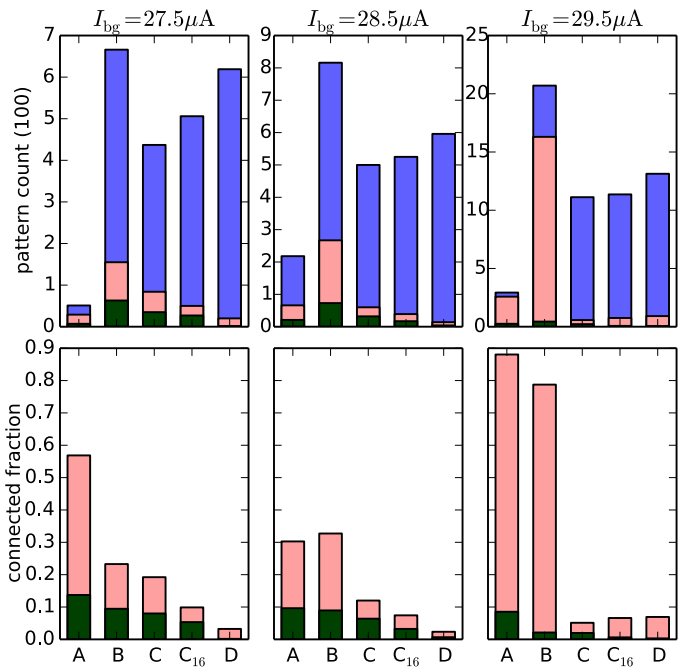}
\par\end{centering}

\protect\caption{(Color online) The number of repeating spike patterns identified from
spiking data of 8 bursts for the five networks in Figs. \ref{fig:Different-network-structures}
and \ref{fig:Average-path-lengths} for different levels of background
current. Blue (full) bars in top panels are the total numbers of identified
patterns from the code in \cite{sun2010selforganization}. Pink (light
gray) bars are for the number of identified patterns that form a connected
subnetwork in the original nominal networks used to generate the spiking
data. Green (dark) bars are for the number of patterns whose neurons
are actually connected following the same sequence of spike times
in each pattern. The upper panels show the number of patterns in unit
of hundreds while the lower panels show the fractions of connected
patterns relative to the total number detected.\label{fig:pattern-stat}}
\end{figure}
 Figure~\ref{fig:pattern-stat} shows the results of the statistics
of the identified repeating spike patterns for these two properties.
A remarkable feature of Fig.~\ref{fig:pattern-stat} is that the
number of detected repeating spike patterns are high in networks with
narrow degree distributions (C, ${\rm C}_{16}$, and D). That is:
repeating spike patterns are more frequent in network with narrow
degree distribution. Unfortunately, most of the detected patterns
are consisted of nodes that do not form a connected subnetwork in
the original network. These repeating spike patterns will give erroneous
results if they are used for the reconstruction of the original network.

Furthermore, for those patterns with nodes forming a connected subnetwork,
only a small fraction of them are actually linked sequentially following
the spike-time orders in the patterns. On the other hand, although
there are far less repeating spike patterns detected for the scale-free
network A, a much more significant fraction of these patterns are
actually consisted of nodes forming connected subnetwork in the original
nominal networks; especially for higher background current $I_{{\rm bg}}=29.5$
$\mu$A. It can be seen from Fig.~\ref{fig:pattern-stat} that the
connected fraction of detected patterns for the scale-free network
seems to be smallest at $I_{{\rm bg}}=28.5$ $\mu$A. This is mainly
due to the peaking of spurious patterns since the \emph{number} of
connected patterns decreases monotonically with $I_{{\rm bg}}$.

\begin{figure}
\includegraphics[width=0.24\textwidth]{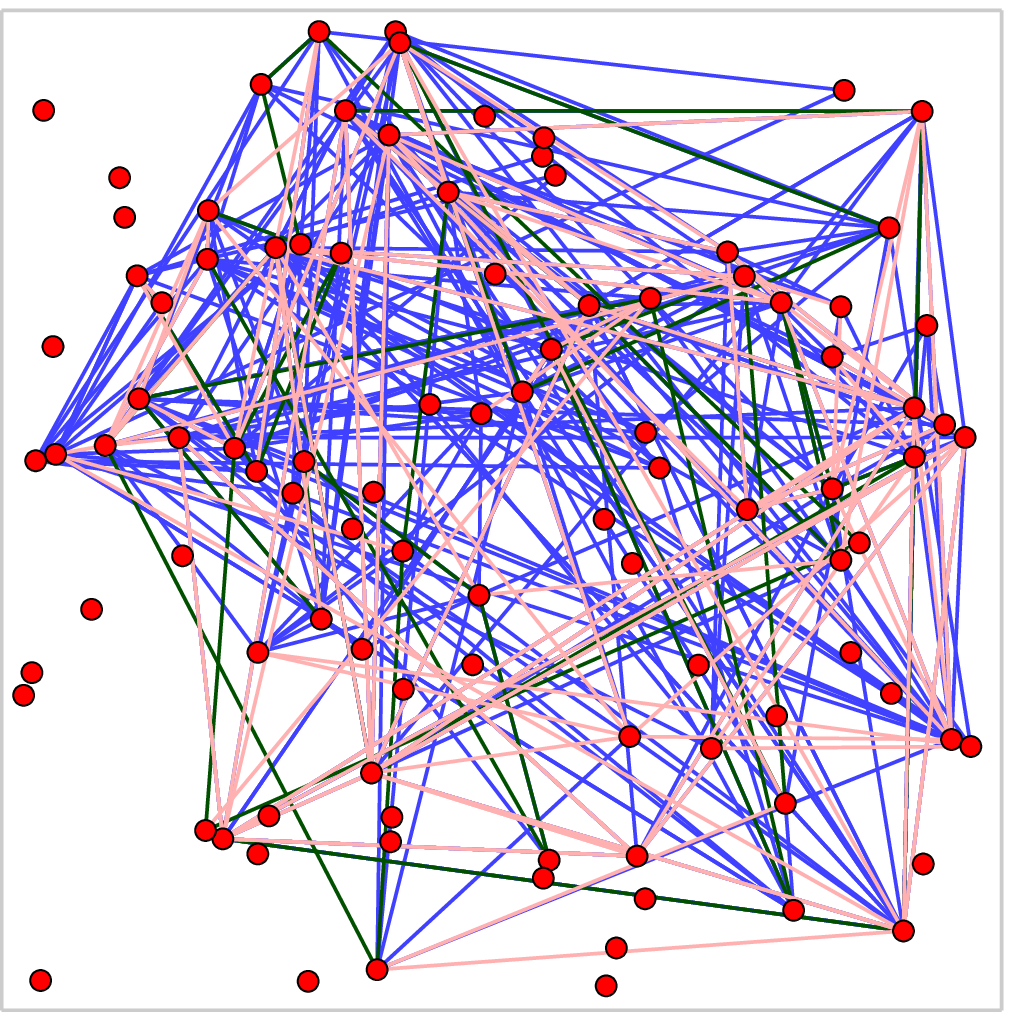}\includegraphics[width=0.24\textwidth]{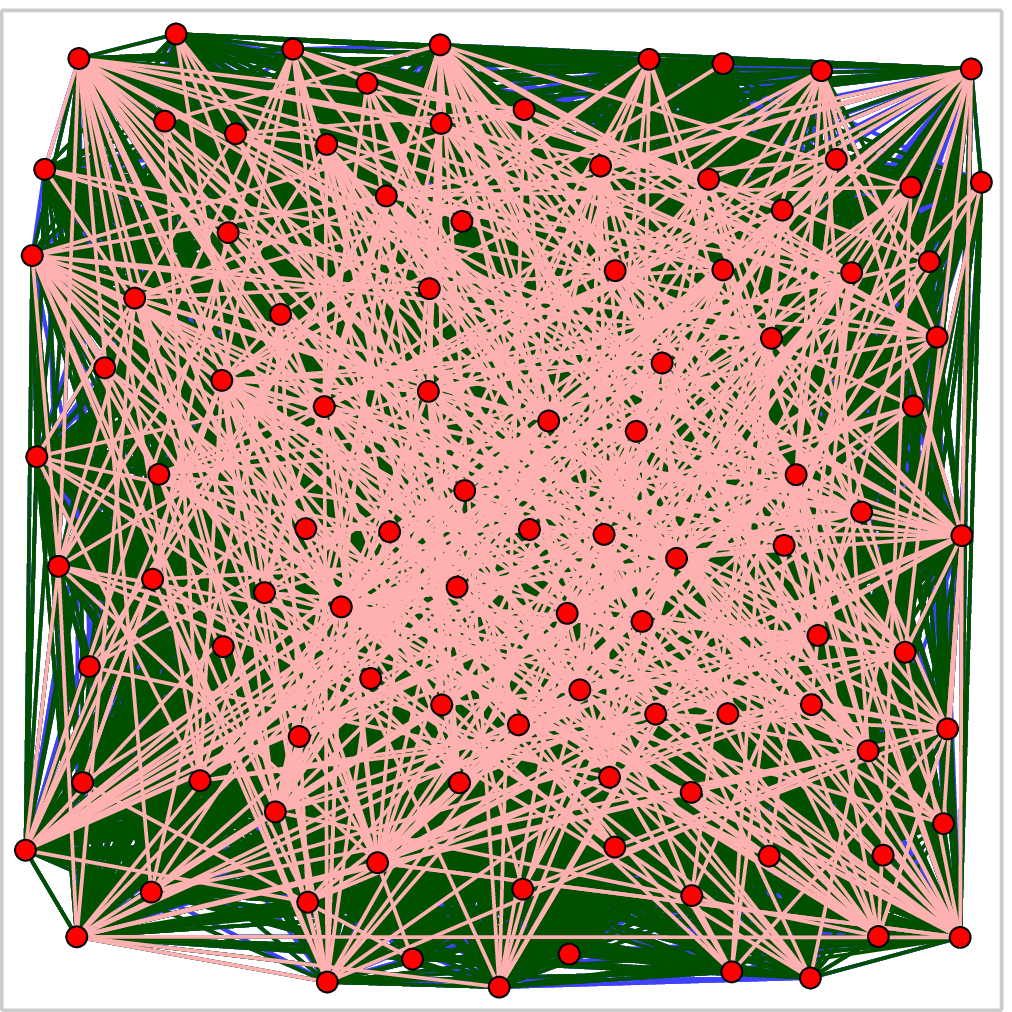}

\includegraphics[width=0.24\textwidth]{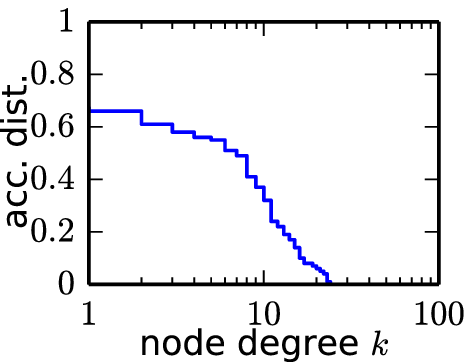}\includegraphics[width=0.24\textwidth]{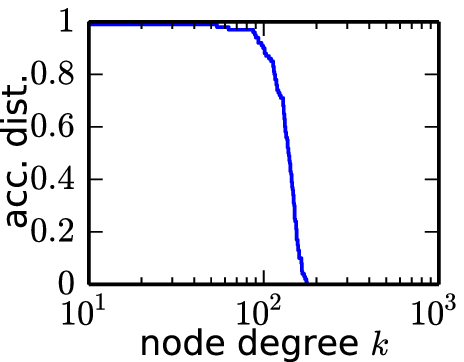}

\protect\caption{(Color online) Structures of the reconstructed networks (upper panels)
from dynamics of scale-free network A (left) and random network D
(right) as well as their corresponding degree distributions (lower
panels). The links colored in pink (light gray), green (dark), and
blue (medium gray) are respectively constructed from patterns of length
$\leq8$, $>32$, and otherwise.\label{fig:reconstructed_networks}}
\end{figure}
Based on the repeating spike patterns identified, we can follow the
method in \cite{sun2010selforganization} to obtain the reconstructed
functional network for the four cases as shown in Fig.~\ref{fig:reconstructed_networks}.
It can be seen from the figure that the physical connection of the
functional network is not only quite different from the physical network,
but global property such as the degree distribution of the functional
network is also not similar to the physical network.

\section{Discussions}

From the results obtained above, it is clear that functional networks
reconstructed from repeating spike patterns can be quite different
from the original physical network. The major reason is that branching
in the propagation of neural activity is ignored in our method of
reconstruction \cite{sun2010selforganization} with the ``fire together
wire together'' heuristic rule. In our simulations, branching is
actually quite important in all the networks studied here. This has
actually been considered by Izhikevich \cite{izhikevich2006polychronization}
in the so called polychronization of spiking activity. Therefore,
we have adopted a relaxed criteria to assess the correlation between
the repeating spike patterns and the nominal network structures based
on whether the nodes participated in a repeating spike pattern form
a connected subnetwork. This is the minimal requirement for a spanning
tree to exist on the nominal network that connects all the firing
nodes of a repeating spike pattern. We note that, this does not imply
that patterns fail to be connected are completely fallacious since
some key nodes could have been left out due to numerical error in
the template-matching algorithm. In the current study, we do not further
investigate how the nominal network structure underlying a repeating
spike pattern can be recovered from the spike pattern itself.

 Also, we have shown (Fig.~\ref{fig:pattern-stat}) that the effectiveness
of using the repeating spike patterns to reconstruct the underlying
nominal network structure is strongly dependent on the topology of
the underlying network and the operating conditions when the spikes
are generated. One major difference between a scale-free network and
a random network is the existence of well-connected hubs. These hubs
can play important roles in relaying information and instigate or
modulate activities of the system \cite{bak1999hownature}. A possible
reason for the higher fractions of connected repeating spike patterns
found for the scale-free A and intermediate B networks is that the
high degree of connectivity for these hubs helps the repeating spike
patterns they participated in form connected subnetworks of the systems.
\begin{figure}
\begin{centering}
\includegraphics[width=0.5\textwidth]{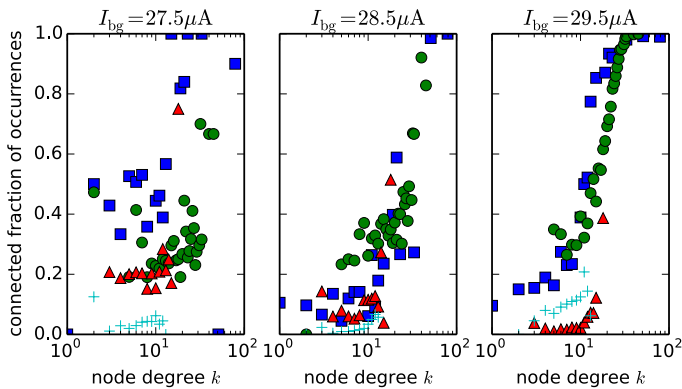}
\par\end{centering}

\protect\caption{(Color online) Mean fractions of node occurrences in detected patterns
that are connected as functions of node degrees. The four different
networks are represented by different symbols: A: squares, B: circles,
C: triangles, D: pluses.\label{fig:Fraction-of-detected}}
\end{figure}
 To verify this hypothesis, we plot in Fig.~\ref{fig:Fraction-of-detected}
the mean fractions of a node's occurrences in repeating spike patterns
that form connected subnetworks versus the degree of the node. It
is evident from Fig.~\ref{fig:Fraction-of-detected} that patterns
of a node with higher degree are more likely to be connected and the
highly connected hub nodes in networks A and B are indeed responsible
for the high fractions of connected patterns. For larger background
current, there exists a degree threshold $k_{c}\approx30$ above which,
all patterns participated by such a node are connected. Such threshold
is not clearly evident for lower background current where the synaptic
weights are stronger to compensate the lower excitability of neurons.

In our simulations, we find that some of the repeating spike patterns
can be embedded into other repeating spike patterns; especially those
with short sequence lengths. That is, these repeating spike patterns
are subsequences of other repeating spike patterns. These repeating
spike patterns are defined as core patterns in \cite{sun2010selforganization}
and are thought to be more relevant to the underlying structure of
the network. However, we find that in a typically simulations, less
than 10\% of all the detected repeating spike patterns are core patterns
by the criteria in \cite{sun2010selforganization} and only a small
number of nodes in the network will take part in these core patterns;
especially for network C and D \cite{SuppMatt}. If we apply our analysis
by using core patterns, nodes participated in the core patterns do
have a higher fraction to be connected except for the random network
D, and the break down to node degree also shows sharper transitions
than those shown in Fig.~\ref{fig:Fraction-of-detected} for core
patterns \cite{SuppMatt}. It is still not clear whether the repeating
spike patterns or the core patterns are better suited for network
reconstruction. Presumably, the core patterns are controlled by the
\textquotedblleft core nodes\textquotedblright{} which are likely
the hubs in the network while almost all nodes in the network can
participate in the repeating spike patterns as long as excitation
(noise) in the system is strong enough. In this aspect, the repeating
spike patterns should be more relevant in providing information for
the reconstruction of the network topology.

The effects of hubs on functional reconstruction can also be seen
in the intermediate networks. The network B in Fig.~\ref{fig:Different-network-structures}
represents an intermediate case between scale-free network with power-law
tail in the degree distribution and a more uniformly connected network
with Gaussian degree distribution. While there are still a number
of hub neurons in such a network, their degrees are limited comparing
to the case of scale-free network. Consequently, unlike the scale-free
network, all neurons can stop firing during the resting period of
its dynamics. We see clear reverberatory bursts in the mid range of
background current $I_{{\rm bg}}=28.0\sim29.0$ $\mu$A \cite{SuppMatt}.
Applying the template-matching method, the spiking dynamics of such
a network produces the largest numbers of repeating spike patterns
in all conditions studied here. Similar to scale-free network A, the
number of connected patterns increases consistently with an increase
of background current. Still, only a small fraction of the patterns
are actually connected sequentially following their spike times. The
picture emerges from our simulations is that although the functional
network reconstructed from repeating spike patterns can be quite different
from the underlying physical network and even the degree distribution
recovered can be quite different, the repeating-spike-pattern method
can be effective in identifying hubs when hubs exist. Since the form
of reverberations are quite different for networks with and without
hubs (Fig~\ref{fig:Comparison-of-bursting}), the form of reverberations
together with the reconstruction by repeating spike patterns might
provide a reliable method to detect hubs in the neuronal cultures.

In the current setup, the characteristics of detected repeating spike
patterns in the small-world network ${\rm C}_{16}$ seem to well interpolate
that of the geometrically constrained network C and random network
D. The low fractions of connected patterns can be correlated with
the absence of hubs in these networks following the discussion above.
There is a steady increase in the number of detected patterns as a
geometrically constrained network becomes more random by rewiring.
However, the number of these patterns that form connected subnetworks
of the underlying nominal network actually becomes less as more links
are rewired. While the rewired long links introduce more possibilities
for activity propagation, hence more patterns, the neighborhood, cluster
structure for a geometrically constrained network can make it more
likely for an identified repeating spike pattern to be connected.
We note that for the networks of 100 neurons in the current study,
the difference in path lengths is very limited for typical consideration
of small-world networks. While our results provide a hint on what
can be expected, to properly address the small-world effect in repeating
spike patterns of simulated neuron networks, significant computing
resources need to be invested to consider networks of higher orders
of magnitudes in sizes.

Since functional networks reconstructed from repeating spike patterns
are governed by both the topology and dynamics of the network and
their excitations are the outcome of collective dynamics of the system,
perhaps the functional network structure is more relevant to the concept
of cell assembly. In fact, Hebb \cite{hebb1949theorganization} had
used the phrase ``functional unit'' when he referred to connections
in a cell assembly. In this sense, the repeating spike patterns provide
a convenient way to look at the complexity of the system. While their
correspondence to the underlying structure of the network can be complicated
by various factors as we have shown in the current study, they are
perhaps more relevant to the functional dynamics of the network that
might serve to fulfill certain biological purposes. The situation
is somewhat similar to relating the gene expression in a cell with
the primary structure of its DNA sequence. Recently, a novel method
\cite{pires2014modeling} based on covariance and time delay is used
to extract the functional connection between different astrocytes
during the propagation of ${\rm Ca}^{2+}$ waves induced by an external
stimulation. However, this method probably cannot be applied to the
case of spontaneous reverberations studied here because in order for
the method in \cite{pires2014modeling} to work, special constrains
are introduced for the network reconstruction. These constrains are
related to single source of stimulation and the directional nature
of the ${\rm Ca}^{2+}$ wave propagation (time delay). In the case
of spontaneous reverberations, both of these features are absent.
In this work, we studied only spontaneous network dynamics. Similar
methods of pattern matching have been used successfully in the identification
of functional unit by using evoked activities. In our current study,
the excitations for the networks are provided by noise. If some of
the nodes in the nominal networks are used as inputs, our simulation
method can also be used to study the repeating spike patterns in evoked
activities.
\begin{acknowledgments}
This work has been supported by the NSC of ROC under the grant nos.
NSC 100-2923-M-001-008-MY3, 101-2112-M-008-004-MY3, and the NCTS of
Taiwan.
\end{acknowledgments}

\begin{thebibliography}{31}
\bibitem [{rat()}]{ratio}%
  \BibitemOpen
  \bibinfo {note} {In experiments with developing neuronal cultures, we find
  that the restitution ratio is of the order of 100 and becomes smaller as the
  cultures mature. In the simulation model, the bursting time and the resting
  time can be adjusted by changing $\tau_l$ and $\tau_s$ respectively. The
  ratio 3 is chosen for convenience and speed of simulation. One could have
  used a ratio of 4 to 1 or 2 to 1. The generic properties of the data
  described here will not be different.}\BibitemShut {Stop}%
\end{thebibliography}
\end{document}